# GMA Instrumentation of the Athena Framework using NetLogger

Dan Gunter, Wim Lavrijsen, David Quarrie, Brian L. Tierney, Craig E. Tull
*Lawrence Berkeley National Laboratory, Berkeley, CA 94720, USA*

Grid applications are, by their nature, wide-area distributed applications. This WAN aspect of Grid applications makes the use of conventional monitoring and instrumentation tools (such as top, gprof, LSF Monitor, etc) impractical for verification that the application is running correctly and efficiently. To be effective, monitoring data must be "end-to-end", meaning that all components between the Grid application endpoints must be monitored. Instrumented applications can generate a large amount of monitoring data, so typically the instrumentation is off by default. For jobs running on a Grid, there needs to be a general mechanism to remotely activate the instrumentation in running jobs. The NetLogger Toolkit Activation Service provides this mechanism.

To demonstrate this, we have instrumented the ATLAS Athena Framework with NetLogger to generate monitoring events. We then use a GMA-based activation service to control NetLogger's trigger mechanism. The NetLogger trigger mechanism allows one to easily start, stop, or change the logging level of a running program by modifying a trigger file. We present here details of the design of the NetLogger implementation of the GMA-based activation service and the instrumentation service for Athena. We also describe how this activation service allows us to non-intrusively collect and visualize the ATLAS Athena Framework moni-

## 1.0 Introduction

Grid monitoring is the measurement and publication of the state of a Grid component at a particular point in time. To be effective, monitoring must be "end-to-end", meaning that all components between the application endpoints must be monitored. This includes software (e.g., applications, services, middleware, operating systems), end-host hardware (e.g., CPUs, disks, memory, network interface), and networks (e.g., routers, switches, or end-to-end paths).

Monitoring is required for a number of purposes, including status checking, troubleshooting, performance tuning, and debugging. For example, assume a Grid job which normally takes 15 minutes to complete has been running for two hours but has not yet completed. Determining what, if anything, is wrong is difficult and requires a great deal of monitoring data. Is the job still running or did one of the software components crash? Is the network congested? Is the CPU loaded? Is there a disk problem? Was a software library containing a bug installed somewhere? Monitoring provides the information to help track down the current status of the job and locate any problems.

A complete end-to-end Grid monitoring system has several components, including:

*Instrumentation*: Instrumentation is the process of putting probes into software or hardware to measure the state of a hardware component, such as a host, disk, network, or a software component, such as operating system, middleware, or application. These probes are often called Sensors.

*Monitoring Event Publication*: Consumers of monitoring event data need to locate appropriate monitoring event providers. Standard schemas, publication mechanisms, and access policies for monitoring event data are required.

*Event Archives*: Archived monitoring event data is critical for performance analysis and tuning, as well as for accounting. Historical data can be used to establish a baseline upon which to compare current performance.

*Sensor Management*: As Grids environments become bigger and more complex, there are more components to monitor and manage, and the amount of monitoring data produced by this effort can quickly become overwhelming. Some components require constant monitoring, while others only are monitored on demand. A mechanism for *activating* sensors on demand is required.

In this paper we describe a Grid activation service which is designed to address the problem of starting, stopping, and changing the level of instrumentation data from running Grid processes and its use in instrumenting the ATLAS software framework Athena. The activation service and control of the instrumentation level is done in a manner that is completely transparent to the Athena application and does not affect its performance. The activation service is designed to work in a cluster environment, and be efficient and scalable. We have run the instrumented Athena application on the PDSF cluster for High Energy and Nuclear Physics at the National Energy Research Scientific Computing (NERSC) facility at Lawrence Berkeley Laboratory. The activation service then collects the instrumentation results, and forwards them to all interested consumers of this data at any distributed client location specified.

The activation service is built using components from NetLogger [15] and pyGMA. pyGMA is our implementation of the Global Grid Forum Grid Monitoring Architecture [1]. NetLogger is used to instrument Grid applications and services, and includes the ability to change the logging-level on the fly by periodically examining a configuration file [11]. The NetLogger binary data format provides an extremely efficient, light-weight transport mechanism for the monitoring data. pyGMA provide an easy to use,

SOAP-based framework for control messages. pyGMA also provide a standard publish-subscribe API for Grid monitoring event publication.

## 1.1 Instrumented Athena

Consider the problem of developing, tuning, and running the Atlas Athena Framework [1] in a Grid Environment. Athena is an object-oriented framework designed to provide a common infrastructure and environment for simulation, filtering, reconstruction and analysis applications high-energy physics experiments. The first step is to insert instrumentation code to ensure the program is operating as expected. This can be done using an instrumentation package such as NetLogger, and instrumentation code should be added to generate timestamped monitoring events before and after CPU intensive tasks, and before and after all disk and network I/O, as is explained in [11].

Once the application is debugged and tested, it is ready for production use. Other monitoring services now become important. The level of instrumentation required for the debugging scenario above can easily generate thousands of monitoring events per second. Clearly one does not need or want this level of monitoring activated all the time, so some type of monitoring *activation service* is needed so that a user can turn instrumentation on and off in a running service.

Next, it is useful to establish a performance baseline for this service, and store this information in the monitoring event archive. System information such as processor type and speed, OS version, CPU load, disk load, and network load data should be collected during the baseline test runs. The monitoring event publication service is needed to locate the sensors and initiate a subscription for the resulting monitoring data. Several tests are then run, sending complete application instrumentation (for clients, servers, and middleware), host, and network monitoring data to the archive. A more detailed example is given in [10].

The components required for this scenario are shown in Figure 1. Athena jobs are running on nodes of one or more compute clusters. The user contacts a monitoring data registry to locate the activation service that is managing the instrumentation level and producing monitoring data for these Athena jobs. The user requests of the activation service that the instrumentation level be increased from the default level (e.g., just error conditions) to a higher level (e.g., full performance trace). The user then subscribes for the instrumentation data, telling the activation service to send the data both to the monitoring archive and back to the user. The activation service collects the data from each of the cluster nodes, and forwards it to both the user and to the monitoring archive.

More details on each of these components are in Section 5.0, below.

## 2.0 Related Work

There are many monitoring systems out there, such as the Condor project's Hawkeye [9], which have publish/subscribe interfaces and some sort of filtering capabilities. Like Hawkeye, these systems are not concerned with application instrumentation and its low thresholds for intrusiveness and lack of direct control mechanisms. Conversely, kernel instrumentation packages such as MAGNet [6] are extremely efficient, but often assume the data can be stored in memory until program exit, and also may require kernel modifications. There are also automatic and semi-automatic application-level instrumentation systems such as Paradyn [12], which are efficient but have simple models for delivering the results and often are specialized for a particular programming model (e.g., parallel programming codes). Although all these systems share some goals with the Activation Service, none have the particular focus on efficient,

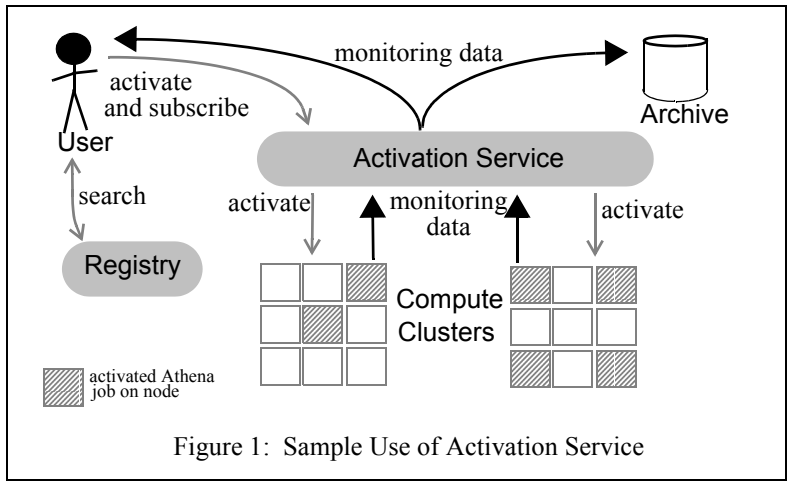

Figure 1: Sample Use of Activation Service

general-purpose application instrumentation in a wide-area distributed setting.

The Open Grid Services Architecture (OGSA) [7] incorporates at a fundamental level much of the functionality required to implement a Grid monitoring service. Any OGSA Grid service can have associated with it arbitrary service data elements (SDEs): named and typed data objects that can be both queried and subscribed to by other parties. The Open Grid Services Infrastructure (OGSI) [17] specification provides specific behaviors for the *notification* interface outlined in the OGSA.

## 2.1 Grid Monitoring Architecture

In 1999 a working group was formed within the Global Grid Forum with the goal of defining a scalable architecture for Grid monitoring. This group has produced both a set of requirements for Grid monitoring, and a high-level specification for a Grid Monitoring Architecture (GMA).

In GMA, the basic unit of monitoring data is called an event. An event is a named, timestamped, structure that may contain one or more items of data. This data may relate to one or more resources such as memory or network usage, or be application-specific data like the amount of time it took to multiply two matrices. The component that makes the event data available is called a producer, and a component that requests or accepts event data is called a consumer. A directory service is used to publish what event data is available and which producer to contact to request it. These components are shown in Figure 3.

GMA supports both a subscription model and a request/response model. In the former case, event data is streamed over a persistent "channel" that is established with an initial request. In the latter case, one item of event data is returned per request. The unique feature of GMA is that performance monitoring data travels directly from the producers of the data to the consumers of the data.

A producer and consumer can be combined to make what is called a producer/consumer pipe. This can be used, for example, to filter or aggregate data. For example, a consumer might collect event data from several producers, and then use that data to generate a new derived event data type, which is then made available to other consumers. More elaborate filtering, forwarding, and caching behaviors could be implemented by connecting multiple consumer/producer pipes.

A number of groups have are now developing monitoring services based on the GMA architecture, such as R-GMA [3] (Relational GMA, so-called because it uses a relational model for all data, organizes data about Grid entities in tables), REMOS [4], and TOPOMON [2].

The OGSA notification service is very similar to GMA. The OGSA interface specifies a notification *source* and *sink*, which are very similar to a *producer* and *consumer* in the GMA. However the current OGSI specification does not provide an unsubscribe operation, or specify a subscription language, and the notification sink requires that all messages be XML. Exposing the Activation Producer as an OGSI notification source would, therefore, consist of uniquely naming the Activation Service's subscription language, adding an unsubscribe operation, and using the existing "locator" element (contained in a subscription) to indicate the NetLogger endpoint. Further integration with NetLogger as a recognized transport protocol for the events will require more support from the OGSI specification.

## 3.0 NetLogger

At Lawrence Berkeley National Lab we have developed the *NetLogger Toolkit* [16], which is designed to monitor, under actual operating conditions, the behavior of all the elements of the application-to-application communication path in order to determine exactly where time is spent within a complex system. Using NetLogger, distributed application components are modified to produce timestamped logs of "interesting" events at all the critical points of the distributed system. Events from each component are correlated, which allows one to characterize the performance of all aspects of the system and network in detail.

The NetLogger library is very efficient, and also easy to use. Using the binary format, NetLogger can serialize on the order of half a million events per second [7]. In order to instrument an application to produce event logs, the application developer inserts calls to the NetLogger API at all the critical points in the code, then links the application with the NetLogger library. This facility is currently available in several languages: Java, C, C++, Python, and Perl. The API has been kept as simple as possible, while still providing automatic timestamping of events and logging to either memory, a local file, syslog, a remote host. Sample Python NetLogger API usage is shown in Figure 2. As is shown in this example, "interesting" events in the code (such as I/O or processing) are typically wrapped with NetLogger write() calls that generate user-defined start and end instrumentation events

The NetLogger ASCII format consists of a whitespace-separated list of "field=value" pairs. Required fields are DATE, HOST, PROG, NL.EVENT and LVL; these can be followed by any number of user-defined

Sample NetLogger event:

```
DATE=20000330112320.957943 HOST=dpss1.lbl.gov PROG=testProg
LVL=Usage NL.EVNT=WriteData SEND.SZ=49332
```

```
log = netlogger.open("x-netlog://loghost.lbl.gov","w")
done = 0
while not done:
    log.write(0,"EVENT_START","TEST.SIZE=%d",size)
    # perform the task to be monitored
    done = do_something(data,size)
    log.write(0,"EVENT_END")
```

Figure 2: Sample NetLogger Usage

fields. The field NL.EVNT contains a unique identifier for the event being logged.

The sample NetLogger event says that the program testprog on host dpss1.lbl.gov performed a WriteData event with a send size of 49,322 on March 30, 2000 at 11:23 (and some seconds) in the morning.

The NetLogger binary format is much faster, but harder for third-party tools to use. NetLogger includes tools for converting between the ASCII and binary formats.

### 3.1 Grid Event Transport

Typically, instrumentation systems only address the problem of extracting the data and writing it to memory or local disk. In a Grid environment, it is just as important to have a robust, efficient means for transporting the instrumentation data beyond that "first hop", to one or more consumers, each of whom may be interested in a different subset of the same instrumentation data. A transport to accomplish this needs to overcome several challenges. Opening connections across the WAN is expensive, so the transport should be able to stream an arbitrary amount of data across a network connection. Temporary network failures in the Grid are the rule, not the exception, so the transport must be reliable in the face of, e.g., broken TCP connections. Because pauses to write out instrumentation are rarely tolerable to the application, data should be buffered before every potential bottleneck (e.g., before any WAN hop).

Finally, delivering a different subset of the data to different consumers requires applying filters on the data at intermediate nodes. To make this feasible on a Grid scale, we believe that the encoding rules and underlying data model should be part of the transport. The encoding offered should be efficient: the component being monitored cannot be perturbed, and intermediaries should be able to apply filters or analyze the data at close to generated rates. In order to help make the encoding efficient, and also to simplify the task of creating and processing the data, the data model should be minimal (i.e., not relational or XML-Infoset).

NetLogger has been designed to answer these requirements of a Grid transport. It has the following features:

- Efficient streaming. NetLogger improves streaming efficiency by buffering all writes for up to 64K or 1 second.
- Reliability. The *write* API allows the user to specify a "backup" file. If a TCP connection fails, the log data is saved to the backup and, optionally, automatically sent over once that connection comes back up again.
- Buffering. The Activation Service directs all logging to local disk, and then reads from these disk buffers in order to forward the data to consumers.
- Efficient encoding. NetLogger has an efficient binary, and very readable ASCII, format. The NetLogger API's can transparently handle both.
- Minimal data model. Each logged item, or "event", is a timestamped set of typed name/value pairs.

In the heterogeneous environment of the Grid, sources and sinks of information may have to dynamically choose an acceptable transport for their required information. To allow NetLogger to participate as one possible transport, we have written a WSDL description of NetLogger.

### 4.0 pyGMA

The pyGMA [13], for "Python GMA", is our implementation of the Grid Monitoring Architecture (GMA) Producer, Consumer, and Registry. It implements Web-Services SOAP interfaces in Python, a high-level object-oriented language. It uses SOAP to aid with serialization and deserialization of messages. Using the pyGMA, only a small amount of Python code would be needed to subscribe to a Producer (e.g., the Activation Producer) for events, directing the results to be transported using NetLogger or query a Producer for one or more events (returned directly in XML).

### 5.0 Activation Service components

The Activation Service has three main components: the *Activation Node*, the *Activation Producer*, the *Activation Manager*. When multiple activation services are deployed, a fourth component, the *Registry*, is also needed. These components are deployed as shown in Figure 3, with one activation node per host, one activation producer and manager per logical host group (e.g., a cluster), and only one (distributed) registry per "Grid".

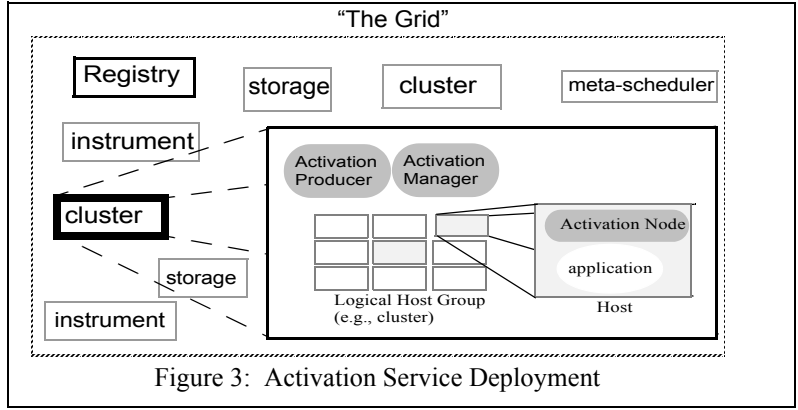

Figure 3: Activation Service Deployment

## 5.1 Activation Node

The Activation Node is responsible for getting the desired logging level from the Activation Manager and communicating this level to the appropriate NetLogger-instrumented programs. It is also responsible for forwarding the instrumentation and monitoring data from these programs to the Activation Producer.

Applications that wish to be activatable must use the NetLogger trigger API, which causes NetLogger to automatically periodically check a "trigger" file for updates to the logging level or destination. In addition, the trigger API will create a small file describing the NetLoggerized application; the Activation Node scans for these files in order to figure out which activatable applications are running on a host. To get the user-specified logging level, the Activation Node polls the Activation Manager using the pyGMA "query" operation, matches the results with the list of known NetLogger-instrumented programs running on this host, and modifies the NetLogger "trigger" file accordingly. Although it would be possible for the Activation Node to tell applications to log directly to the Activation Producer, this may cause delays if the Activation Producer becomes overloaded. Therefore, the Activation Node always "triggers" logging to a temporary file on local disk, and forwards the monitoring data asynchronously to the Activation Producer.

## 5.2 Activation Producer

The Activation Producer receives pyGMA subscriptions from consumers. As mentioned above, it also receives NetLogger instrumentation data from the Activation Node. The main task of the Activation Producer, then, is to match incoming instrumentation data with the subscriptions. In order to do this efficiently, the monitoring data is multiplexed, demultiplexed, and filtered by a NetLogger "pipe", part of the standard NetLogger library. In addition to using the efficient NetLogger encoding, by performing these functions inside the NetLogger library we also minimize copying of the monitoring data. Subscriptions are transformed into NetLogger "filters", which are added to the pipe, as illustrated in Figure 4.

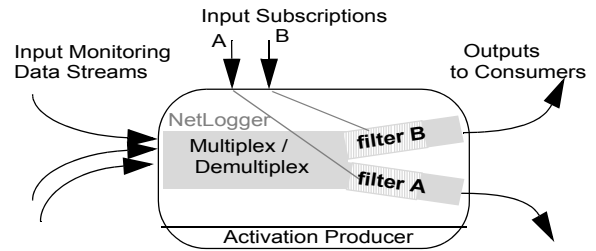

Figure 4: NetLogger in the Activation Producer

Rather than try to build or borrow an expressive but complex filter language such as [ref], we devised a simple method that would handle the most common use cases. NetLogger filters operate on one item of monitoring data at a time, testing to see if that item matches any single *expression*. An expression is a list of (name, operator, value) tuples. For example, a query that matches all "Start" or "End" monitoring events for program "Athena" at a logging level less than or equal to two would be:

NL.EVNT="Start" **and** PROG="Athena" **and** LVL <= 2
               **or**
NL.EVNT="End" and PROG="Athena" and LVL <= 2

This matches the NetLogger event:

    DATE=20030529235002.185091 NL.EVNT=Start HOST=127.0.0.1 PROG=Athena LVL=1

but does not match either of the NetLogger events:

    DATE=20030529235007.518600 NL.EVNT=Middle HOST=127.0.0.1 PROG=Athena LVL=1
    DATE=20030529235007.518600 NL.EVNT=End HOST=127.0.0.1 PROG=Athena LVL=3

Due to the simplicity of the filter language, the implementation is straightforward and efficient. This is important because, as described in Section 5.2, filtering is used extensively by the Activation Producer. Performance results for the NetLogger filtering API are shown in Section 7.2.

### 5.3 Activation Manager

The Activation Manager keeps track of the logging level for a given NetLogger-instrumented application. If the application is logging at level 3, then only log messages of level 0..3 will be produced; a logging level of -1 means "off". NetLogger instrumentation associates a log level with each piece of monitoring data, so an attempt to write monitoring data whose level is above the current logging level results in a no-op. This means that reducing the logging level is an easy and efficient way to reduce the overhead of instrumentation.

The Activation Manager is polled periodically (e.g.: every 5 seconds) by each Activation Node for its current list of "activations". One issue with this design is that with a large number of hosts (e.g., 500) and a small poll interval (e.g., 5 seconds), the request parsing can cause a high load on the Activation Manager host. This potential load is one reason that the architecture separates the Activation Manager from the Activation Producer.

### 5.4 Registry

Consumers who want to subscribe for monitoring data can search the Registry for the appropriate instance of the Activation Service. A typical search would be "find me the Activation Service associated with MyApplication on Cluster A or Cluster B". The Registry will return one or more Activation Service endpoints, and then the user can proceed to subscribe for the data, activate the logging, or both. The Registry could also be used to locate other GMA Producers with the same or related monitoring data, such as a monitoring data archive.

Because there is nothing about the Registry that is specific to the Activation Service, we have not yet attempted an implementation. Existing projects, such as the MDS and R-GMA Registry, should serve admirably. It should be noted that we also have not needed a Registry up to this point, as all experiments have been run on a known cluster with a known associated Activation Service..

### 6.0 Athena Framework Overview

The ATLAS software framework, Athena[1], is an object-oriented framework based upon the GAUDI component architecture originally developed within the LHCb experiment. Athena is designed to provide a common infrastructure and environment for simulation, filtering, reconstruction and analysis applications for high energy physics experiments at the Large Hadron Collider (LHC) at CERN. These experiments are expected to run for many years and therefore changes in software requirements and in the technologies used to build software have to be taken into account by developing flexible and adaptable software that can withstand these changes and which can be easily maintained over the long timescales involved.

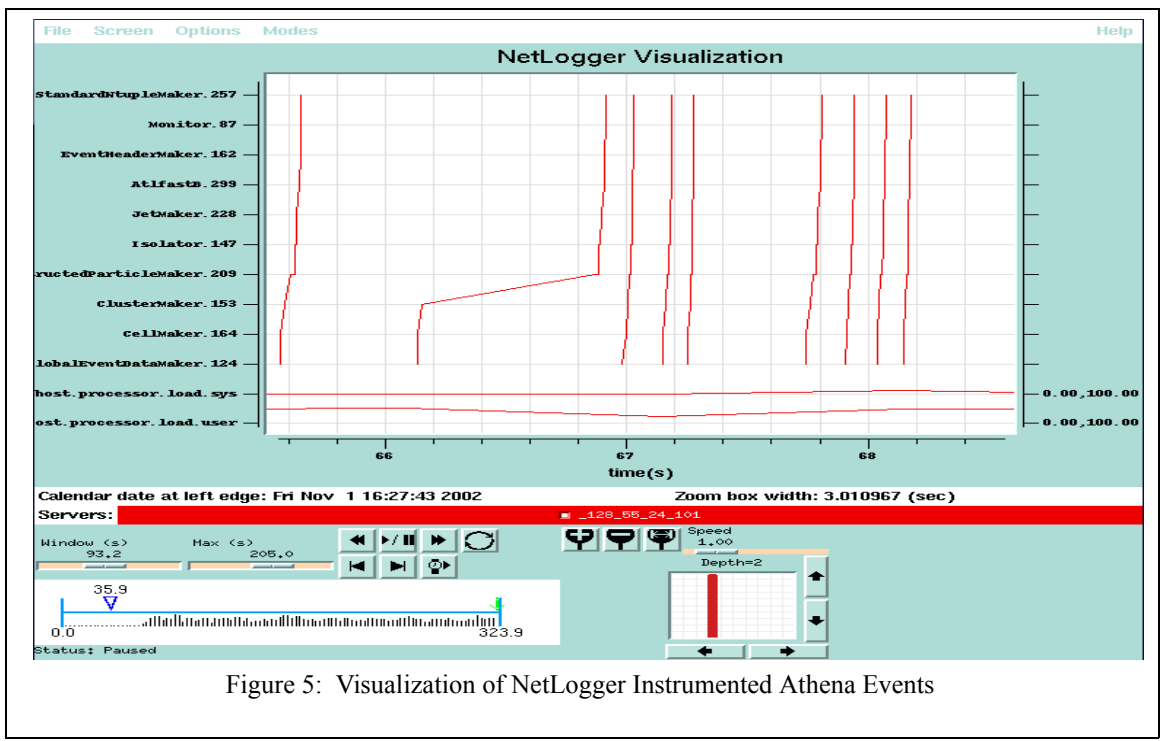

Figure 5: Visualization of NetLogger Instrumented Athena Events

Motivated by the tremendous compute and storage resources required by LHC reconstruction, simulation, and analysis, LHC scientists are at the forefront of user of the Grid for these tasks.

User written code in Athena are encapsulated in C++ objects called Algorithms. These Algorithms' lifecycles and execution control are handled by the generic Athena framework.

Framework services are supplied as pluggable components whose lifecycles are also controlled by the framework, but whose execution is typically a result of Algorithm invokation.

Athena integration with Grid middleware includes NetLogger instrumentation at the entry point, and some within, each Algorithm. A graph of the instrumentation results gathered from an Athena run is shown in Figure 5. Note that the CPU load can be easily graphed side-by-side with instrumentation results.

## 7.0 Results

In this section, we present results from using the Activation Service with an instrumented Athena Framework running on multiple nodes on a cluster. We focus on how the system scales as we increase the number of scheduled application instances (i.e.: number of cluster nodes used) and the number of "consumers" subscribing to the instrumentation data. Both simple and relatively complex subscriptions are employed, and a detailed (40 events/sec) logging level is activated. The next section describes the experimental setup, and subsequent sections present the results.

## 7.1 Experimental Setup

The experimental setup is shown in Figure 6. We ran Athena jobs on a queue in the NERSC "PDSF" cluster (http://pdsf.nersc.gov) in Oakland, CA, which has over 200 compute nodes available at any given time. The NetLogger instrumentation was sent to an Activation Producer, labeled *hostB*, at LBNL (about 15 km away), and the activaton level was set and queried at an Activation Manager, labeled *hostA*, on the same subnet. Subscriptions (generated on a laptop) told the Activation Producer to send monitoring data to two remote hosts, one at Oak Ridge National Laboratory and one at the University of Pittsburgh Supercomputing Center. Five streams of monitoring data were sent to each consumer host. All hosts were 400 MHz or higher Pentium systems running Linux; for more details see Appendix A.

In order to assess the performance of the entire system, various components were instrumented with NetLogger,
and these logs were sent to yet another host on the LBNL subnet.

During each test, we measured the CPU load on the cluster nodes and on the Activation Producer and Activation Manager, and also the latency for each event between the time it was generated and the time it arrived at a Consumer.

## 7.2 Filter Performance

To better understand the performance characteristics of the Activation Service as a whole, we also tested the NetLogger "filter API" in isolation. Good performance here is crucial, as the filtering event rate provides an upper bound on the throughput of the Activation Producer. There are two independent variables which affect the event rate: filter complexity, and the proportion of events which 'pass' the filter. The more complex the filter, the longer it takes to evaluate each event. The more events which 'pass' the filter, the longer NetLogger spends performing I/O.

To evaluate the trade-off, we ran tests where the filter complexity varied from 0 to 40 comparisons in steps of 4, and the proportion of events that 'passed' varied from 0% to 100% in steps of 10%. The events used were similar to those in the Athena instrumentation. Results were logged from a host with the same configuration as *hostB.lbl.gov*, to the remote consumer in Pittsburgh (*host.psc.edu*).

Measuring the throughput versus the two independent variables produced the co-plot shown in Figure 6. For each scatterplot, events per second are on the Y axis, and the filter complexity is on the X axis. From left to right and bottom to top, increasing percentages of events "passed" the filter, i.e. were written to the destination.

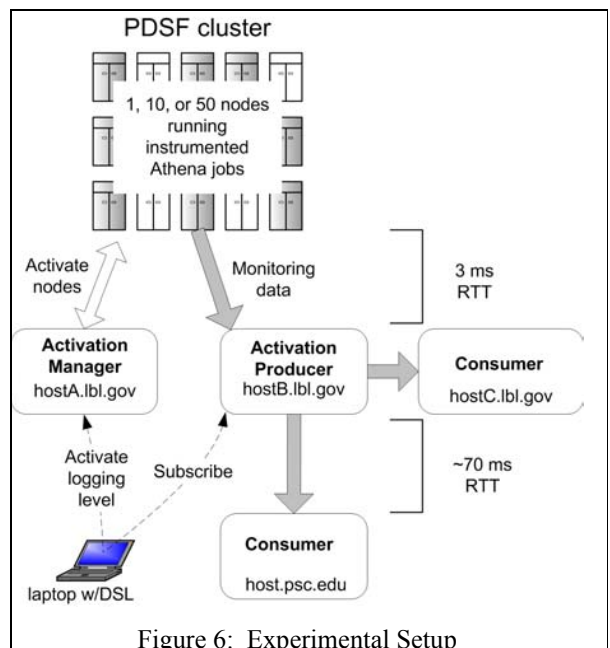

Figure 6: Experimental Setup

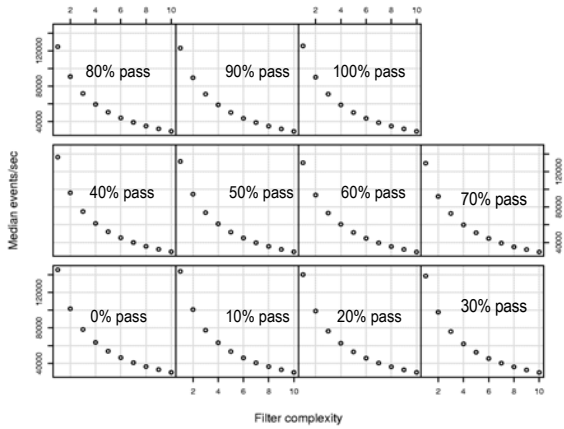

Figure 7: Filter performance results

Clearly, the filter complexity is the main effect on performance. Between the simplest and most complex filters, there is roughly a factor of five slow-down, whereas the 100% pass filter is only about five to ten percent slower than the 0% pass filter.

However, even at reasonably high complexity, the filter performance is good. For example, when the filter has 20 expressions (complexity 5), on average it can write roughly 50,000 events per second. This is still faster than the raw speed of many less efficient logging libraries, such as *log4j* [8]. It should be noted that because the host used was relatively slow (400MHz) and filtering is CPU-intensive, most current hardware would have even better performance.

### 7.3 Scalability

The primary scalability question that we wished to address is: how many consumers and producers (instrumented jobs) can be handled by a single Activation Producer and Activation Manager? Secondarily, how does the complexity of the subscription affect these quantities? We measured the scalability of the system by comparing the average time for an event to travel from the instrumented job to the consumer as the number of producers, consumers, and event rate increased.

In our tests, we "pre-subscribed" to the Activation Producer with either a simple (4-comparison) or complex (20-comparison) filter expression, for each of 10 consumers. Then we submitted roughly 5-minute jobs on PDSF. We measured the event latencies for each event. The median values are graphed in Figure 7.

Before analyzing these results, some fixed sources of latency (particular to this implementation) should be mentioned. First, the forwarding of events from each instrumented job occurs in short bursts separated by 5 seconds. Second, each NetLogger Consumer and Producer uses buffers with a a 1-second timeout to increase streaming efficiency. So, because each event is written, forwarded, read, written, and read again (see Figure 6) -- the average fixed latency overhead is $((1 + 5 + 1 + 1 + 1) \div 2) = 4.5$ seconds.

From 1 to 20 producers, the median event latency varied between 4.7 and 6.2 seconds. Neither consumer location or filter location seem to affect the values. Therefore, it seems that these latencies are random variation just above the fixed latency discussed above. This means that a single Activation Producer scaled to 20 producers at 40 events per second with a complex filter to

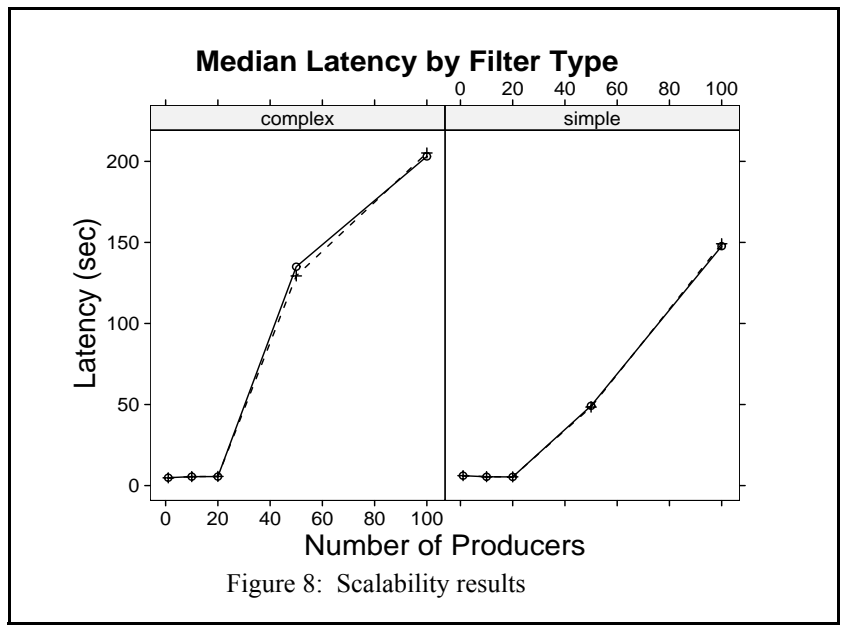

Figure 8: Scalability results

10 consumers, i.e. an aggregate output rate of 8,000 events per second.

From 50 and 100 producers, we see linear increases in latency for both complex and simple filtering. As expected, the complex filter is slower than the simple filter. Again, consumer location does not affect latency.

Tests with 1 consumer, as opposed to 10, showed that latency was near the minimum, even with 100 producers. The reason for this is that the filtering overhead is incurred per-consumer, and therefore 1/10 the number of consumers incur 1/10 as much overhead. This means that Activation Producer performance with 100 producers and 1 consumer is only a little worse than the performance with 10 producers and 10 consumers.

## 8.0 Conclusions

In this paper we described a Grid *activation service* which is designed to address the problem of starting, stopping, and changing the level of instrumentation data from running Grid processes. We have shown that a single Activation Producer node is efficient enough to handle an aggregate throughput of 8,000 monitoring events per second. In order to use the Activation Producer, the NetLogger instrumentation in Athena did not need to be modified at all, and only a single component (the Activation Node) needed to be added to the PDSF filesystem. The query language, although simple, provided sufficient flexibility for current needs. Overall, we believe that the Activation Services's flexible, distributed architecture will prove to be a useful building block for a comprehensive Grid monitoring and troubleshooting system.

The instrumentation with NetLogger of the Athena framework was simple and provided monitoring information which was informative both about the behavior of the Athena Algorithms, and of the GMA activation service.

An abstract interface for a GMA service in the Athena framework has been developed as well as a concrete implementation of this interface using NetLogger. This service can be invoked automatically through the standard Auditors mechanism in Athena which is triggered before and after selected Algorithms.

Further work in the Athena GMA service is needed to allow full instrumentation in a non-intrusive manner. Other concrete implementations of the service can invoke different instrumentation libraries without change to the users' Algorithms or the framework itself.

## 9.0 Acknowledgments

This work was supported by the Office of Science. High Energy Physics, U.S. Department of Energy under Contract No. DE-AC03-76SF00098.

This work was supported by the Director, Office of Science. Office of Advanced Scientific Computing Research. Mathematical, Information, and Computational Sciences Division under U.S. Department of Energy Contract No. DE-AC03-76SF00098.

See the disclaimer at http://www-library.lbl.gov/disclaimer.

This is report no. LBNL-52977.